\documentclass[preprint,amsmath,amssymb,aps]{revtex4-1}

\usepackage{graphicx}
\usepackage{dcolumn}
\usepackage{bm}
\usepackage{epsfig}
\usepackage{epstopdf}
\usepackage{subfig}
\usepackage{caption}
\usepackage{enumerate}
\usepackage{comment}
\usepackage{physics}
\usepackage{amsmath}
\usepackage{dsfont}
\usepackage{slashed}
\usepackage{ulem}
\usepackage{color}
\usepackage{hyperref}
\usepackage [autostyle, english = american]{csquotes}
\MakeOuterQuote{"}
\begin{document}

\begin{flushright}
LFTC-25-11 / 105
\end{flushright}

\title{Ward-Takahashi identity in the light-front formalism for a bound state of fermions}

 \author{Deepesh Bhamre}
  \email{deepeshbhamre7@gmail.com}
 \author{J. P. B. C. de Melo}
  \email{joao.mello@cruzeirodosul.edu.br} 
  \affiliation{Laborat\'{o}rio de F\'{i}sica Te\'{o}rica e Computacional (LFTC), Universidade Cruzeiro do Sul / Universidade Cidade de S\~{a}o Paulo, 01506-000, S\~{a}o Paulo, Brazil}

 \date{\today}

\begin{abstract}
	We investigate the Ward-Takahashi identity at one-loop in the light-front (LF) formalism for a bound state of fermions. We consider a spinless bound state made up of two fermions in which the Ward-Takahashi identity is satisfied in the covariant formulation. Considering the same system in the light-front formalism, we investigate the proof of Ward-Takahashi identity by integrating the light-front energy component through the identification of the relevant ranges of the longitudinal LF momentum. We elucidate that the pair production diagram plays a crucial role in establishing the Ward-Takahashi identity. We also point out the necessity of taking into account the corresponding zero modes for truly establishing the Ward-Takahashi identity in the LF formalism.

\begin{description}

\item[Keywords]
{Ward-Takahashi identity, Light-Front formalism, electromagnetic current, zero modes}
%\item[PACS numbers]
%{}

\end{description}

\end{abstract}

\maketitle

\section{Introduction}\label{sec:intro}

%%%%  \section*{The Problem of Ward-Takahashi Identities on the Light-Front}
Light-front quantum field theory (LFQFT) is an inherently suitable framework for modeling composite systems such as mesons or baryons 
\cite{Brodsky1998}. The Fock amplitudes associated with the eigenstates
 of the light-front Hamiltonian effectively capture the intricate hadron structure that stems from the fundamental interactions of quantum chromodynamics (QCD).

The phenomenological success of this particular approach in describing hadron structure is noteworthy.
Its results are comparable to those obtained using a variety of other well-established methods, including,
Schwinger-Dyson methods \cite{Roberts1994,Roberts2000}, QCD sum rules
\cite{Braguta2004,Aliev2004,Aliev2009}, anti-de Sitter (AdS)/QCD frameworks \cite{Brodsky2007},
effective field theory (EFT)~\cite{Bijnens1998}, covariant light-front dynamics 
\cite{Karmanov1996,Leao2023}, and also point-form quantum mechanics~\cite{Biernat2014}. 

Light-front quantum field theory presents distinct advantages,
 particularly in its formalism, where the vacuum is trivial and the kinematical group
  includes Lorentz transformations \cite{Brodsky1998}. This simplifies the calculation of bound states 
significantly compared to traditional formalisms. 

However, LFQFT is not without challenges. A critical issue is the loss of covariance in certain physical processes 
\cite{deMelo1997,deMelo1999,Bakker2002,Choi2004}.
 To ensure the full covariance of the electromagnetic current, it is
  necessary to go beyond the simple valence component and incorporate nonvalence
   contributions (often called zero modes) into the electromagnetic current's matrix elements~\cite{deMelo1997,deMelo1999,deMelo2012,demelo2018,Meijian2019}.

	The problem of covariance breaking in the light-front (LF) formalism has received considerable attention in the relevant literature. This issue arises because the chosen quantization surface ($x^+ = x^0 + x^3 = 0$) does not preserve manifest rotational symmetry, a subgroup of the full Lorentz group.
	
	In a recent publication \cite{Zhang2025}, specifically concerning the Yukawa model for the nucleon, the problematic inclusion
	 of zero modes ($k^+ = 0$ modes, and nonvalence terms, higher Fock-sector contributions, often referred to as pair terms) was discussed. The dependence of the obtained results on the adopted reference frames 
	was explored, highlighting the necessity of these extra contributions to recover the expected
	Lorentz invariance and frame independence of physical observables, such as form factors.

	In the LF approach, the matrix elements of the current operator $\hat{J}^\mu$ are typically calculated using the overlap of light-front wave functions. For the full covariance to hold, the results derived from different current components, $\langle P' | \hat{J}^+ | P \rangle$ and $\langle P' | \hat{J}^- | P \rangle$, must be consistent. The necessity to include zero modes or nonvalence contributions,
	is often a consequence of the truncation of the Fock space or the specific component of the current employed. The authors in \cite{Zhang2025} (and references therein) investigate how the proper treatment of these terms is essential for the cancellation of spurious frame dependence, a clear signature of the restoration 
	of Poincaré symmetry~\cite{deMelo1998,deMelo1999}.

	Ward-Takahashi identities express how the symmetries inherent in a quantum field theory, such as gauge symmetry, govern the connections between its physical quantities, specifically its correlation functions and amplitudes. These constraints are so fundamental that they must be maintained throughout the technical steps of regularization and renormalization. 
	
	The Ward-Takahashi (WT) identities  \cite{Ward1950,Takahashi1957}, stemming from local $U(1)$
	 gauge invariance, are fundamental results in conventional quantum field theory (QFT). It is therefore remarkable that they have not been adequately addressed within the framework of  light-front  field theory for 
	 fermions case. The primary difficulty lies in the fact that the established formal proofs 
	\cite{Ward1950,Takahashi1957,Naus1998,Marinho2007,Marinho2008}, which crucially depend on equal-time canonical commutation relations, are inapplicable to the light-front. 

Light-front dynamics is inherently a constrained system, 
and thus precludes direct canonical quantization. This noncanonical structure introduces complications, 
particularly in fermionic theories, where certain components 
of the spinor fields are not independent degrees of freedom; 
they are subject to constraint equations, and these are often referred to as bad components. 
	These bad components necessarily enter into the definition of the  electromagnetic operators, 
		 subsequently leading to modified, potentially nonstandard currents.	
	Despite the fact that the simplest Ward-Takahashi identity connects the electromagnetic
	 vertex and the propagator, the primary unanswered question involves
	  the integral relationship tying the current to the vertex.
	
Some research has focused on the significance of the  pair-creation and annihilation terms  (often referred to as simply pair terms) within the context of the  bosonic model in references \cite{Naus1998,Martinovic2018}. It was previously demonstrated that these pair terms remain relevant, even when considering the zero light-front momentum limit,
$p^{+} \to 0$, and are  essential for ensuring a covariant electromagnetic
 current \cite{deMelo1998}.

	The current investigation significantly advances the initial results by exploring a more generalized scope (or a broader class of systems/parameters). Central to this expanded analysis is a detailed examination of the WT identities as they apply to the fermionic sector of the theory. Crucially, the analysis reaffirms the persistent requirement to incorporate nonvalence contributions—those originating outside the simplest quasiparticle picture, to accurately model the electromagnetic current. 
	
	This substantiates the inadequacy of purely valence-only approximations for a complete description with the light-front approach.
	
In the present work, with a one-loop current in the Yukawa model, the only dynamical fields in the loop integrals are massive fermions. Consequently, the zero-mode contributions, which are critical for infrared structure and for phenomena like confinement in certain frameworks, stem solely from these particles. Within this framework, the fermion mass $m$ provides a suppression of the fermionic zero-mode effects, thereby yielding a more subdued infrared structure compared to gauge-theoretic scenarios. However, in non-Abelian gauge theories like QCD, the dominant infrared dynamics, including the generation of a confining potential, is primarily driven by the gauge boson zero modes. The gluon, in actuality massless in the Lagrangian, provides unsuppressed contributions in the infrared regime. In the case of the Abelian quantum electrodynamics, the photon zero modes can lead to significant long-range effects, which are absent in purely Yukawa systems.

The present article is organized as follows. In Sec. \ref{sec:WT_Covariant}, we provide a proof for the Ward-Takahashi identity at one loop in the covariant formulation for the given system. It is followed by the proof in the LF formulation in Sec. \ref{sec:WT_LF}. In Sec. \ref{subsec:Self_energy}, the LF expression for self-energy is obtained by integrating out the energy component in the loop momentum. The contribution of zero modes to this expression is explained in the process. Further, the LF expression for the electromagnetic current is similarly obtained in Sec. \ref{subsec:Current} and the LF Ward-Takahashi identity is established. A brief qualitative discussion about electromagnetic form factor in the context of the issue at hand is given in Sec. \ref{subsec:em_form_factor}. Section \ref{sec:concl} contains a few concluding remarks. The conventions used throughout the article and a few useful identities are provided in Appendix \ref{app:conventions}, and a part of the algebra from Sec. \ref{subsec:Current} is relegated to Appendix \ref{app:integrals_algebra}.

\section{Ward-Takahashi identity in the covariant formulation}\label{sec:WT_Covariant}

\begin{figure}[h!]
\centering
%\subfloat[]
{\includegraphics[scale=0.65]{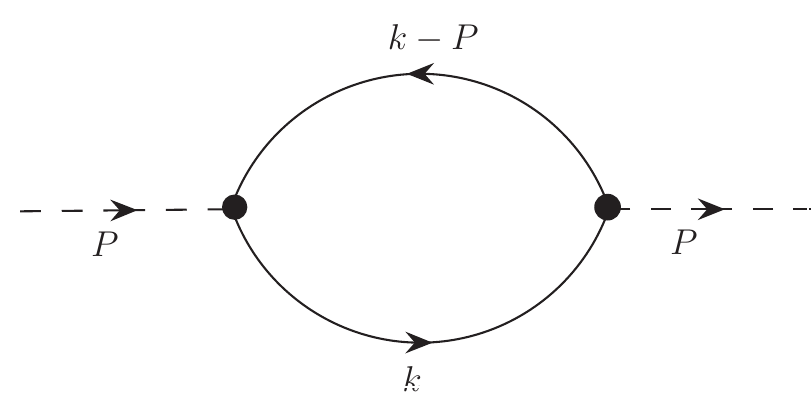}}
\caption{Self-energy of a spin-$0$ bound-state of a fermion-antifermion pair}
\label{fig:self_en}
\end{figure}

Consider a spin-$0$ bound-state made up of a fermion-antifermion pair, for example a pion, given by a $\gamma^{5}$ type of an interaction vertex. The self-energy diagram for such a state is shown in Fig. \ref{fig:self_en}.

The expression for the self-energy of this bound state is given in the equal-time covariant formulation by 
\begin{equation}\label{eq:self_en}
\Sigma(P) = \int\frac{d^{4}k}{(2\pi)^4}\frac{Tr[(\slashed{k}+m)\gamma^{5}(\slashed{P'}-\slashed{k}-m)\gamma^{5}]}{(k^{2}-m^{2}+i\epsilon)[(P-k)^{2}-m^{2}+i\epsilon]}.
\end{equation}
The trace appearing due to the fermion loop equals
\begin{equation}\label{eq:tr_self_en}
Tr[(\slashed{k}+m)\gamma^{5}(\slashed{P}-\slashed{k}-m)\gamma^{5}] = 4(k^{2}-m^{2}-k.P).
\end{equation}
Thus,
\begin{equation}
\Sigma(P) = 4 \int\frac{d^{4}k}{(2\pi)^4}\frac{k^{2}-m^{2}-k.P}{(k^{2}-m^{2}+i\epsilon)[(P-k)^{2}-m^{2}+i\epsilon]}.
\end{equation}

Hence, the difference in the self-energy diagrams with external momenta $P$ and $P'$ of the bound states is,
\begin{equation}\label{eq:self_en_diff}
\begin{split}
&\Sigma(P)-\Sigma(P')\\
& = 4 \int\frac{d^{4}k}{(2\pi)^4}\bigg[\frac{k^{2}-m^{2}-k.P}{(k^{2}-m^{2}+i\epsilon)[(P-k)^{2}-m^{2}+i\epsilon]}-\frac{k^{2}-m^{2}-k.P'}{(k^{2}-m^{2}+i\epsilon)[(P'-k)^{2}-m^{2}+i\epsilon]}\bigg]
\end{split}.
\end{equation}

\begin{figure}[h!]
\centering
%\subfloat[]
{\includegraphics[scale=0.65]{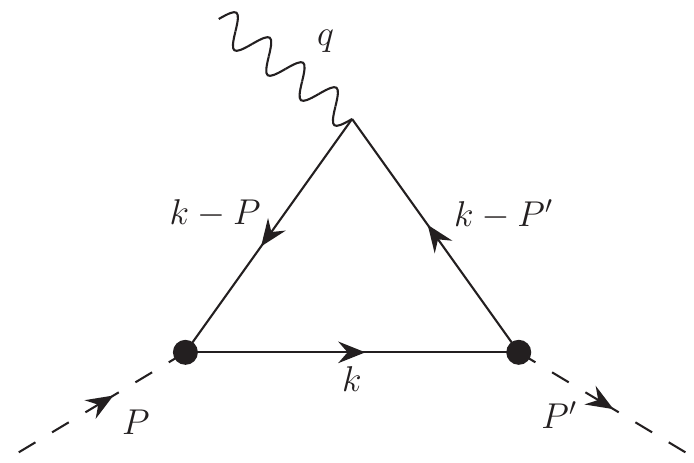}}
\caption{Triangle diagram for the electromagnetic current}
\label{fig:triangle}
\end{figure}

Now consider the electromagnetic current corresponding to this bound state. It is represented by the Feynman triangle diagram shown in Fig. \ref{fig:triangle}.
The covariant current in equal-time formulation is given by
\begin{equation}\label{eq:current}
J^{\mu}(P,P') = \int\frac{d^{4}k}{(2\pi)^4}\frac{Tr[(\slashed{k}+m)\gamma^{5}(\slashed{P'}-\slashed{k}-m)\gamma^{\mu}(\slashed{P}-\slashed{k}-m)\gamma^{5}]}{(k^{2}-m^{2}+i\epsilon)[(P'-k)^{2}-m^{2}+i\epsilon][(P-k)^{2}-m^{2}+i\epsilon]},
\end{equation}
where the electromagnetic charge is taken to be $e=1$. The trace appearing in the current due to the fermion loop equals
\begin{equation}\label{eq:tr_current}
\begin{split}
&Tr[(\slashed{k}+m)\gamma^{5}(\slashed{P'}-\slashed{k}-m)\gamma^{\mu}(\slashed{P}-\slashed{k}-m)\gamma^{5}]\\
& = 4[(k^{2}-m^{2})(P'+P-k)^{\mu}+(P'.P)k^{\mu}-(k.P')P^{\mu}-(k.P)P'^{\mu}].
\end{split}
\end{equation}

With the photon momentum $q_{\mu} = (P'-P)_{\mu} = (P'-k)_{\mu}-(P-k)_{\mu}$, we get
\begin{equation}\label{eq:q.J}
\begin{split}
&q_{\mu}J^{\mu}(P,P')\\
& = 4\int\frac{d^{4}k}{(2\pi)^4}\frac{[(P'-k)_{\mu}-(P-k)_{\mu}][(k^{2}-m^{2})(P'+P-k)^{\mu}+(P'.P)k^{\mu}-(k.P')P^{\mu}-(k.P)P'^{\mu}]}{(k^{2}-m^{2}+i\epsilon)[(P'-k)^{2}-m^{2}+i\epsilon][(P-k)^{2}-m^{2}+i\epsilon]}.
\end{split}
\end{equation}
The numerator of the integrand in Eq. (\ref{eq:q.J}) is 
\begin{equation}\label{eq:num_of_q.J}
\begin{split}
&[(P'-k)_{\mu}-(P-k)_{\mu}][(k^{2}-m^{2})(P'+P-k)^{\mu}+(P'.P)k^{\mu}-(k.P')P^{\mu}-(k.P)P'^{\mu}]\\
& = (P'-k)^{2}(k^{2}-m^{2})+(P'-k)_{\mu}[(k^{2}-m^{2})P^{\mu}+(P'.P)k^{\mu}-(k.P')P^{\mu}-(k.P)P'^{\mu}]\\&\phantom{ = } - (P-k)^{2}(k^{2}-m^{2})-(P'-k)_{\mu}[(k^{2}-m^{2})P'^{\mu}+(P'.P)k^{\mu}-(k.P')P^{\mu}-(k.P)P'^{\mu}]\\
& = [(P'-k)^{2}-m^{2}](k^{2}-m^{2}-k\cdot P)- [(P-k)^{2}-m^{2}](k^{2}-m^{2}-k\cdot P'),
\end{split}
\end{equation}
where $[(k\cdot P)+(k\cdot P')]k^{\mu}$ is added and subtracted to arrive at the last equation. Substituting in Eq. (\ref{eq:q.J}), we get
\begin{equation}
\begin{split}
&q_{\mu}J^{\mu}(P,P')\\
& = 4\int\frac{d^{4}k}{(2\pi)^4}\frac{[(P'-k)^{2}-m^{2}](k^{2}-m^{2}-k\cdot P)- [(P-k)^{2}-m^{2}](k^{2}-m^{2}-k\cdot P')}{(k^{2}-m^{2}+i\epsilon)[(P'-k)^{2}-m^{2}+i\epsilon][(P-k)^{2}-m^{2}+i\epsilon]}\\
& = 4\int\frac{d^{4}k}{(2\pi)^4}\bigg[\frac{k^{2}-m^{2}-k\cdot P}{(k^{2}-m^{2}+i\epsilon)[(P-k)^{2}-m^{2}+i\epsilon]} - \frac{k^{2}-m^{2}-k\cdot P'}{(k^{2}-m^{2}+i\epsilon)[(P'-k)^{2}-m^{2}+i\epsilon]}\bigg]\\
& = \Sigma(P)-\Sigma(P'),
\end{split}
\end{equation}
as can be seen from Eq. (\ref{eq:self_en_diff}).

Thus, the Ward-Takahashi identity, given by $q_{\mu}J^{\mu}(P,P') = \Sigma(P)-\Sigma(P')$ at the one-loop level for a single incoming and outgoing particle, is proven to be satisfied in the covariant formulation.

\section{Ward-Takahashi identity in the LF formulation}\label{sec:WT_LF}

The light-front formulation is usually implemented in the form a Hamiltonian field theory. Thus, all particles in the theory are on shell. We now establish the Ward-Takahashi identity at one-loop for the system in Sec.\ref{sec:WT_Covariant} i.e. a fermion-antifermion bound-state of spin 0, but in the LF formulation. For this, we consider the 4D loop integral in the covariant theory and integrate out the energy component from the loop integral. This procedure is essentially equivalent to calculating the time-ordered graphs of the theory in old-fashioned Hamiltonian perturbation theory \citep{equivalencepramana}. While establishing the Ward-Takahashi identity such, one needs to be mindful of the zero modes that arise when the longitudinal LF momentum vanishes. We proceed to show this below.

\subsection{The self-energy}\label{subsec:Self_energy}

The self-energy for the considered bound-state, as given in Sec.\ref{sec:WT_Covariant}, is 
\begin{equation}
\Sigma(P) = \int\frac{d^{4}k}{(2\pi)^4}\frac{Tr[(\slashed{k}+m)\gamma^{5}(\slashed{P}-\slashed{k}-m)\gamma^{5}]}{(k^{2}-m^{2}+i\epsilon)[(P-k)^{2}-m^{2}+i\epsilon]}. \tag{\ref{eq:self_en}}
\end{equation}
In LF coordinates,
\begin{equation}
\begin{split}
\slashed{k} &= \gamma^{+}k^{-} + \gamma^{-}k^{+} - \gamma^{\perp}\cdot k^{\perp}\\
&= \gamma^{+}\bigg(\frac{k_{\perp}^{2}+m^{2}}{2k^{+}}\bigg) + \gamma^{-}k^{+} - \gamma^{\perp}\cdot k^{\perp} + \gamma^{+}\bigg(k^{-}-\frac{k_{\perp}^{2}+m^{2}}{2k^{+}}\bigg)\\
&= \gamma^{+}k^{-}_{\text{on}} + \gamma^{-}k^{+} - \gamma^{\perp}\cdot k^{\perp} + \frac{\gamma^{+}}{2k^{+}}(k^{2}-m^{2}),
\end{split}
\end{equation}
where $k^{-}_{\text{on}}=\frac{k_{\perp}^{2}+m^{2}}{2k^{+}}$ is the on-shell LF energy of a particle with momentum $k^{\mu}=(k^{+}, k^{-}, k^{\perp})$. Thus, a fermion propagator can be split into an ``on-shell" term and an ``off-shell" term as follows 
\begin{equation}\label{eq:split_mom}
\frac{\slashed{k}+m}{k^{2}-m^{2}} = \frac{\slashed{k}_{\text{on}}+m}{k^{2}-m^{2}} + \frac{\gamma^{+}}{2k^{+}},
\end{equation}
the on-shell $4$-momentum being given by $k^{\mu}_{\text{on}} = (k^{+}, k^{-}_{\text{on}}, k^{\perp})$. The ``off-shell'' term is a nonpropagating term that is typically present in the LF formulation. Similarly, the other propagator in the fermion loop is written as
\begin{equation}
\frac{\slashed{P}-\slashed{k}+m}{(P-k)^{2}-m^{2}} = \frac{(\slashed{P}-\slashed{k})_{\text{on}}+m}{(P-k)^{2}-m^{2}} + \frac{\gamma^{+}}{2(P^{+}-k^{+})}.
\end{equation}

Splitting the propagator momenta thus, Eq. (\ref{eq:self_en}) takes the following form in LF coordinates.

\begin{equation}
\begin{split}
\Sigma(P) = \int\frac{d^{2}k_{\perp}dk^{+}dk^{-}}{(2\pi)^4}\bigg[& \frac{Tr[(\slashed{k}_{\text{on}}+m)\gamma^{5}[(\slashed{P}-\slashed{k})_{\text{on}}-m]\gamma^{5}]}{2k^{+}\big(k^{-}-\frac{f_{1}-i\epsilon}{2k^{+}}\big)2(P^{+}-k^{+})\big(P^{-}-k^{-}-\frac{f_{2}-i\epsilon}{2(P^{+}-k^{+})}\big)}\\
& + \frac{Tr[(\slashed{k}_{\text{on}}+m)\gamma^{5}\gamma^{+}\gamma^{5}]}{2k^{+}\big(k^{-}-\frac{f_{1}-i\epsilon}{2k^{+}}\big)2(P^{+}-k^{+})}\\
& + \frac{Tr[\gamma^{+}\gamma^{5}[(\slashed{P}-\slashed{k})_{\text{on}}-m]\gamma^{5}]}{2k^{+}2(P^{+}-k^{+})\big(P^{-}-k^{-}-\frac{f_{2}-i\epsilon}{2(P^{+}-k^{+})}\big)}\\
& + \frac{Tr[\gamma^{+}\gamma^{5}\gamma^{+}\gamma^{5}]}{2k^{+}2(P^{+}-k^{+})}\bigg],
\end{split}
\end{equation}
where $f_{1} = k_{\perp}^{2}+m^{2}$,\\
\phantom{where }$f_{2} = (P_{\perp}-k_{\perp})^{2}+m^{2}$.

Using the properties of gamma matrices $\{\gamma^{5},\gamma^{\mu}\} = 0$ and $(\gamma^{+})^{2} = 0$, it is seen that the last term of the above integrand vanishes. Also, the traces in the remaining terms can be calculated using the properties of gamma matrices, giving\\
$Tr[(\slashed{k}_{\text{on}}+m)\gamma^{5}[(\slashed{P}-\slashed{k})_{\text{on}}-m]\gamma^{5}] = -4(k_{\text{on}}\cdot (P-k)_{\text{on}}+m^{2})$,\\
$Tr[(\slashed{k}_{\text{on}}+m)\gamma^{+}] = 4k^{+}$,\\
$Tr[[(\slashed{P}-\slashed{k})_{\text{on}}-m]\gamma^{+}] = 4(P^{+}-k^{+})$.

Hence, the expression for self-energy in the LF coordinates becomes
\begin{equation}\label{eq:self-en_three_terms}
\begin{split}
\Sigma(P) = -\int\frac{d^{2}k_{\perp}dk^{+}dk^{-}}{(2\pi)^4}\bigg[& \frac{k_{\text{on}}\cdot (P-k)_{\text{on}}+m^{2}}{k^{+}(P^{+}-k^{+})\big(k^{-}-\frac{f_{1}-i\epsilon}{2k^{+}}\big)\big(P^{-}-k^{-}-\frac{f_{2}-i\epsilon}{2(P^{+}-k^{+})}\big)}\\
& + \frac{1}{(P^{+}-k^{+})\big(k^{-}-\frac{f_{1}-i\epsilon}{2k^{+}}\big)} + \frac{1}{k^{+}\big(P^{-}-k^{-}-\frac{f_{2}-i\epsilon}{2(P^{+}-k^{+})}\big)}\bigg].
\end{split}
\end{equation}

As mentioned previously, the $k^{-}$-integration now needs to be done to arrive at the expression for self-energy in the LF Hamiltonian formalism. The above equation can be written as
\begin{equation}\label{eq:sum of I1 to I3}
\Sigma(P) = -\int\frac{d^{2}k_{\perp}dk^{+}}{(2\pi)^4}\big[I_{1}^{-}+I_{2}^{-}+I_{3}^{-}\big],
\end{equation}
where
\begin{equation}\label{eq:I1_self_en}
I_{1}^{-} = \int dk^{-}\frac{k_{\text{on}}\cdot (P-k)_{\text{on}}+m^{2}}{k^{+}(P^{+}-k^{+})\big(k^{-}-\frac{f_{1}-i\epsilon}{2k^{+}}\big)\big(P^{-}-k^{-}-\frac{f_{2}-i\epsilon}{2(P^{+}-k^{+})}\big)},
\end{equation}

\begin{equation}\label{eq:I2_self_en}
I_{2}^{-} = \int dk^{-}\frac{1}{(P^{+}-k^{+})\big(k^{-}-\frac{f_{1}-i\epsilon}{2k^{+}}\big)},
\end{equation}

\begin{equation}\label{eq:I3_self_en}
I_{3}^{-} = \int dk^{-}\frac{1}{k^{+}\big(P^{-}-k^{-}-\frac{f_{2}-i\epsilon}{2(P^{+}-k^{+})}\big)}.
\end{equation}

$I_{1}^{-}$ in Eq. (\ref{eq:I1_self_en}) has poles at $k_{1}^{-} = \frac{f_{1}-i\epsilon}{2k^{+}}$ and $k_{2}^{-} = P^{-}-\frac{f_{2}-i\epsilon}{2(P^{+}-k^{+})}$. Note that there is no $k^{-}$ dependence in the numerator as the $4$-momenta appearing there ($k^{\mu}_{\text{on}}$, $(P-k)^{\mu}_{\text{on}}$) are on-shell momenta. For $k^{+}< 0$, both the poles $k_{1}^{-}$ and $k_{2}^{-}$ lie above the real axis while for $k^{+}> P^{+}$ , both lie below it. Hence, closing the contour on the opposite side of the position of the pole, the integral vanishes for each of these two ranges of the longitudinal momentum. For $0<k^{+}<P^{+}$, $k_{1}^{-}$ lies below the real axis whereas $k_{2}^{-}$ lies above. Closing the contour below, the residue theorem gives
\begin{equation}
I_{1}^{-} = \frac{-2\pi i(k_{\text{on}}\cdot (P-k)_{\text{on}}+m^{2})\theta(k^{+})\theta(P^{+}-k^{+})}{k^{+}(P^{+}-k^{+})\big(P^{-}-\frac{f_{1}}{2k^{+}}-\frac{f_{2}}{2(P^{+}-k^{+})}\big)}.
\end{equation}

Note that in Eq. (\ref{eq:I1_self_en}), there is no end-point singularity present at $k_{1}^{-}\to \infty$ for $k^{+}\to 0$ or at $k_{2}^{-}\to -\infty$ for $(P^{+}-k^{+})\to 0$. This is because there are sufficient powers of the integration variable in the denominator to eliminate the contribution of the semicircular arc of the contour. Hence, there is no contribution to the first term of Eq. (\ref{eq:self-en_three_terms}) in the limit of vanishing longitudinal momentum of either of the fermions in the loop. In other words, there is no zero-mode contribution.

However, this is not the case in Eqs.(\ref{eq:I2_self_en}) and (\ref{eq:I3_self_en}). $I_{2}^{-}$ in Eq. (\ref{eq:I2_self_en}) has a pole at $k_{1}^{-} = \frac{f_{1}-i\epsilon}{2k^{+}}$ which does go to infinity as $k^{+}\to 0$. This is a purely divergent zero-mode term that manifests here as an end-point singularity. In order to correctly account for this pole at infinity, we make a change of variable so as to bring the pole to the origin, regularize it, and eventually take the regulator to zero \citep{equivalencepramana}. Let $k^{-} = \frac{1}{u}$ in Eq. (\ref{eq:I2_self_en}). Then,
\begin{equation}
I_{2}^{-} = \int_{-\infty}^{\infty}du\frac{1}{(P^{+}-k^{+})u\big[1-u\big(\frac{f_{1}-i\epsilon}{2k^{+}}\big)\big]}.
\end{equation}
Regulating the pole, which is now at the origin, by using a small parameter $\delta$ to write $\frac{1}{u} = \frac{1}{2}\big[\frac{1}{u+i\delta}+\frac{1}{u-i\delta}\big] + \mathcal{O}(\delta^{2})$, we get
\begin{equation}
\begin{split}
I_{2}^{-} = & \frac{1}{2}\int_{-\infty}^{\infty}du\frac{1}{(P^{+}-k^{+})(u+i\delta)\big[1-u\big(\frac{f_{1}-i\epsilon}{2k^{+}}\big)\big]}\\
&+ \frac{1}{2}\int_{-\infty}^{\infty}du\frac{1}{(P^{+}-k^{+})(u-i\delta)\big[1-u\big(\frac{f_{1}-i\epsilon}{2k^{+}}\big)\big]}.
\end{split}
\end{equation}
The first integral above has poles at $u_{1}=-i\delta$ and $u_{2}=\frac{2k^{+}}{f_{1}-i\epsilon}$. Thus, the integral is nonvanishing only for $k^{+} > 0$. Similarly, the second integral contributes only through the region $k^{+} < 0$. Applying the residue theorem, and then taking the regulator $\delta$ to zero, we get
\begin{equation}
I_{2}^{-} = \frac{-2\pi i \theta(k^{+})}{2(P^{+}-k^{+})} + \frac{2\pi i \theta(-k^{+})}{2(P^{+}-k^{+})}.
\end{equation}

Another method to account for the end-point singularity in Eq. (\ref{eq:I2_self_en}) is provided in Refs.\citep{bakkerdewitt, equivalencepramana}.

In a similar manner, $I_{3}^{-}$ in Eq. (\ref{eq:I3_self_en}) is evaluated to give
\begin{equation}
I_{3}^{-} = \frac{2\pi i \theta(k^{+}-P^{+})}{2k^{+}} - \frac{2\pi i \theta(P^{+}-k^{+})}{2k^{+}}.
\end{equation}
Substituting the evaluated integrals $I_{1}^{-}$ through $I_{3}^{-}$ in Eq. (\ref{eq:sum of I1 to I3}), we get
\begin{equation}
\begin{split}
\Sigma(P) = i\int\frac{d^{2}k_{\perp}}{(2\pi)^3}\bigg[&\int_{0}^{P^{+}}dk^{+}\frac{k_{\text{on}}\cdot (P-k)_{\text{on}}+m^{2}}{k^{+}(P^{+}-k^{+})\big(P^{-}-\frac{f_{1}}{2k^{+}}-\frac{f_{2}}{2(P^{+}-k^{+})}\big)}\\
& + \int_{0}^{\infty}\frac{dk^{+}}{2(P^{+}-k^{+})} - \int_{-\infty}^{0}\frac{dk^{+}}{2(P^{+}-k^{+})} - \int_{P^{+}}^{\infty}\frac{dk^{+}}{2k^{+}} + \int_{-\infty}^{P^{+}}\frac{dk^{+}}{2k^{+}}\bigg].
\end{split}
\end{equation}
Changing the variable $k^{+}\to P^{+}-k^{+}$ in the second and third terms, it can be easily shown that
\begin{equation}
\begin{split}
\Sigma(P) = i\int\frac{d^{2}k_{\perp}}{(2\pi)^3}\bigg[&\int_{0}^{P^{+}}dk^{+} \frac{k_{\text{on}}\cdot (P-k)_{\text{on}}+m^{2}}{k^{+}(P^{+}-k^{+})\big(P^{-}-\frac{f_{1}}{2k^{+}}-\frac{f_{2}}{2(P^{+}-k^{+})}\big)}\\
& + \int_{-\infty}^{P^{+}}\frac{dk^{+}}{k^{+}} - \int_{P^{+}}^{\infty}\frac{dk^{+}}{k^{+}}\bigg].
\end{split}
\end{equation}
Considering $P'^{+} > P^{+}$, the integration regions of the log-divergent $k^{+}$ integrals combine to give
\begin{equation} \label{eq:WT_self-en}
\begin{split}
\Sigma(P) - \Sigma(P') = i\int\frac{d^{2}k_{\perp}}{(2\pi)^3}\bigg[&\int_{0}^{P^{+}}dk^{+} \frac{k_{\text{on}}\cdot (P-k)_{\text{on}}+m^{2}}{k^{+}(P^{+}-k^{+})\big(P^{-}-\frac{f_{1}}{2k^{+}}-\frac{f_{2}}{2(P^{+}-k^{+})}\big)}\\
& - \int_{0}^{P'^{+}}dk^{+}\frac{k_{\text{on}}\cdot (P'-k)_{\text{on}}+m^{2}}{k^{+}(P'^{+}-k^{+})\big(P'^{-}-\frac{f_{1}}{2k^{+}}-\frac{f_{3}}{2(P'^{+}-k^{+})}\big)}\\
& - 2\int_{P^{+}}^{P'^{+}}\frac{dk^{+}}{k^{+}}\bigg],
\end{split}
\end{equation}
where $f_{3} = (P'_{\perp}-k_{\perp})^{2}+m^{2}$. The last term in Eq. (\ref{eq:WT_self-en}) is the zero-mode contribution to the difference in boson self-energies.

\subsection{The electromagnetic current}\label{subsec:Current}

In LF coordinates, we write Eq. (\ref{eq:q.J}) as
\begin{equation}\label{eq:q.J_LF}
\begin{split}
&q_{\mu}J^{\mu}(P,P')\\
& = 4\int\frac{d^{2}k_{\perp}dk^{+}}{(2\pi)^4}\int dk^{-}\frac{(P'-P)_{\mu}[(k^{2}-m^{2})(P'+P-k)^{\mu}+(P'.P)k^{\mu}-(k.P')P^{\mu}-(k.P)P'^{\mu}]}{(k^{2}-m^{2}+i\epsilon)[(P'-k)^{2}-m^{2}+i\epsilon][(P-k)^{2}-m^{2}+i\epsilon]}.
\end{split}
\end{equation}
The numerator in Eq. (\ref{eq:q.J_LF}) is 
\begin{equation}
\begin{split}
&(P'-P)_{\mu}[(k^{2}-m^{2})(P'+P-k)^{\mu}+(P'.P)k^{\mu}-(k.P')P^{\mu}-(k.P)P'^{\mu}]\\
& = [P'^{2}-P^{2}-(k.P')+(k.P)](k^{2}-m^{2}) + [(k.P')-(k.P)](P'.P)\\
&\phantom{ = } - [(P'.P)-P^{2}](k.P') - [P'^{2}-(P'.P)](k.P)\\
& = [[(P'-k)^{2}-m^{2}] - [(P-k)^{2}-m^{2}] + (k.P') - (k.P)](k^{2}-m^{2})\\
&\phantom{ = } + P^{2}(k.P') - P'^{2}(k.P)\\
& = [(P'-k)^{2}-m^{2}](k^{2}-m^{2}) - [(P-k)^{2}-m^{2}](k^{2}-m^{2})\\
&\phantom{ = } + [(P-k)^{2}-m^{2}](k.P') - [(P'-k)^{2}-m^{2}](k.P).
\end{split}
\end{equation}
Substituting in Eq. (\ref{eq:q.J_LF}) gives
\begin{equation}
\begin{split}
&q_{\mu}J^{\mu}(P,P')\\
& = 4\int\frac{d^{2}k_{\perp}dk^{+}}{(2\pi)^4}\bigg[\int dk^{-}\frac{1}{[(P-k)^{2}-m^{2}+i\epsilon]} - \int dk^{-}\frac{k.P}{(k^{2}-m^{2}+i\epsilon)[(P-k)^{2}-m^{2}+i\epsilon]}\\
&\phantom{ = 4\int\frac{d^{2}k_{\perp}dk^{+}}{(2\pi)^4}\bigg[} - \int dk^{-}\frac{1}{[(P'-k)^{2}-m^{2}+i\epsilon]} + \int dk^{-}\frac{k.P'}{(k^{2}-m^{2}+i\epsilon)[(P'-k)^{2}-m^{2}+i\epsilon]}\bigg].
\end{split}
\end{equation}

The first and third integrals, and the terms proportional to $k^{-}$ in the numerator of the second and fourth integrals contain end-point singularities. Thus, we write the above equation as
\begin{equation}\label{eq:current_minus_int_sum}
q_{\mu}J^{\mu}(P,P') = \int\frac{d^{2}k_{\perp}dk^{+}}{(2\pi)^4} \sum_{i=1}^{6} I^{-}_{i},
\end{equation}
where
\begin{equation}
\begin{split}
I^{-}_1 & = \int dk^{-}\frac{2}{(P^{+}-k^{+})\big(P^{-}-k^{-}-\frac{f_{2}-i\epsilon}{2(P^{+}-k^{+})}\big)}\\
& = \frac{2\pi i \theta(k^{+}-P^{+})}{(P^{+}-k^{+})} - \frac{2\pi i \theta(P^{+}-k^{+})}{(P^{+}-k^{+})},
\end{split}
\end{equation}
\begin{equation}
\begin{split}
I^{-}_2 & = \int dk^{-}\frac{-P^{-}k^{+}+P_{\perp}\cdot k_{\perp}}{k^{+}(P^{+}-k^{+})\big(k^{-}-\frac{f_{1}-i\epsilon}{2k^{+}}\big)\big(P^{-}-k^{-}-\frac{f_{2}-i\epsilon}{2(P^{+}-k^{+})}\big)}\\
& = \frac{2\pi i(P^{-}k^{+}-P_{\perp}\cdot k_{\perp}) \theta(k^{+})\theta(P^{+}-k^{+})}{k^{+}(P^{+}-k^{+})(P^{-}-k^{-}_{\text{on}}-(P-k)^{-}_{\text{on}})},
\end{split}
\end{equation}
\begin{equation}
\begin{split}
I^{-}_3 & = \int dk^{-}\frac{-P^{+}k^{-}}{k^{+}(P^{+}-k^{+})\big(k^{-}-\frac{f_{1}-i\epsilon}{2k^{+}}\big)\big(P^{-}-k^{-}-\frac{f_{2}-i\epsilon}{2(P^{+}-k^{+})}\big)}\\
& = \frac{(2\pi i)P^{+} \theta(-k^{+})}{2k^{+}(P^{+}-k^{+})} + \frac{(2\pi i)P^{+}(P^{-}+k^{-}_{\text{on}}-(P-k)^{-}_{\text{on}}) \theta(k^{+})\theta(P^{+}-k^{+})}{2k^{+}(P^{+}-k^{+})(P^{-}-k^{-}_{\text{on}}-(P-k)^{-}_{\text{on}})}\\
&\phantom{ = } - \frac{(2\pi i)P^{+} \theta(k^{+}-P^{+})}{2k^{+}(P^{+}-k^{+})},
\end{split}
\end{equation}
\begin{equation}
\begin{split}
I^{-}_4 & = \int dk^{-}\frac{-2}{(P'^{+}-k^{+})\big(P'^{-}-k^{-}-\frac{f_{3}-i\epsilon}{2(P'^{+}-k^{+})}\big)}\\
& = \frac{2\pi i \theta(P'^{+}-k^{+})}{(P'^{+}-k^{+})} - \frac{2\pi i \theta(k^{+}-P'^{+})}{(P'^{+}-k^{+})},
\end{split}
\end{equation}
\begin{equation}
\begin{split}
I^{-}_5 & = \int dk^{-}\frac{P'^{-}k^{+}-P'_{\perp}\cdot k_{\perp}}{k^{+}(P'^{+}-k^{+})\big(k^{-}-\frac{f_{1}-i\epsilon}{2k^{+}}\big)\big(P'^{-}-k^{-}-\frac{f_{3}-i\epsilon}{2(P'^{+}-k^{+})}\big)}\\
& = \frac{-2\pi i(P'^{-}k^{+}-P'_{\perp}\cdot k_{\perp}) \theta(k^{+})\theta(P'^{+}-k^{+})}{k^{+}(P'^{+}-k^{+})(P'^{-}-k^{-}_{\text{on}}-(P'-k)^{-}_{\text{on}})},
\end{split}
\end{equation}
\begin{equation}
\begin{split}
I^{-}_6 & = \int dk^{-}\frac{P'^{+}k^{-}}{k^{+}(P'^{+}-k^{+})\big(k^{-}-\frac{f_{1}-i\epsilon}{2k^{+}}\big)\big(P'^{-}-k^{-}-\frac{f_{3}-i\epsilon}{2(P'^{+}-k^{+})}\big)}\\
& = \frac{(2\pi i)P'^{+} \theta(k^{+}-P'^{+})}{2k^{+}(P'^{+}-k^{+})} - \frac{(2\pi i)P'^{+}(P'^{-}+k^{-}_{\text{on}}-(P'-k)^{-}_{\text{on}}) \theta(k^{+})\theta(P'^{+}-k^{+})}{2k^{+}(P'^{+}-k^{+})(P'^{-}-k^{-}_{\text{on}}-(P'-k)^{-}_{\text{on}})}\\
&\phantom{ = } - \frac{(2\pi i)P'^{+} \theta(-k^{+})}{2k^{+}(P'^{+}-k^{+})}.
\end{split}
\end{equation}
Note that all but $I^{-}_2$ and $I^{-}_5$ above contain end-point singularities, providing divergent zero-mode contributions to the electromagnetic current. All integrals $I^{-}_i$ above are evaluated in a similar manner as in Sec.\ref{subsec:Self_energy}. Substituting $I^{-}_1$ through $I^{-}_6$ in Eq. (\ref{eq:current_minus_int_sum}), one can write
\begin{equation}\label{eq:current_plus_int_sum}
q_{\mu}J^{\mu}(P,P') = i\int\frac{d^{2}k_{\perp}}{(2\pi)^3} \sum_{i=1}^{12} I^{+}_{i},
\end{equation}
where
\begin{equation*}
I^{+}_{1} = \int_{P^{+}}^{\infty} \frac{dk^{+}}{P^{+}-k^{+}},\hspace{.4cm} I^{+}_{2} = -\int_{-\infty}^{P^{+}} \frac{dk^{+}}{P^{+}-k^{+}},\hspace{.4cm} I^{+}_{3} = \int_{0}^{P^{+}} dk^{+}\frac{P^{-}k^{+}-P_{\perp}\cdot k_{\perp}}{k^{+}(P^{+}-k^{+})(P^{-}-k^{-}_{\text{on}}-(P-k)^{-}_{\text{on}})},
\end{equation*}
\begin{equation*}
I^{+}_{4} = \int_{-\infty}^{0} dk^{+}\frac{P^{+}}{2k^{+}(P^{+}-k^{+})},\hspace{.4cm} I^{+}_{5} = \int_{0}^{P^{+}} dk^{+}\frac{P^{+}(P^{-}+k^{-}_{\text{on}}-(P-k)^{-}_{\text{on}})}{2k^{+}(P^{+}-k^{+})(P^{-}-k^{-}_{\text{on}}-(P-k)^{-}_{\text{on}})},
\end{equation*}
\begin{equation*}
I^{+}_{6} = -\int_{P^{+}}^{\infty} dk^{+}\frac{P^{+}}{2k^{+}(P^{+}-k^{+})},\hspace{.4cm} I^{+}_{7} = \int_{-\infty}^{P'^{+}} \frac{dk^{+}}{P'^{+}-k^{+}},\hspace{.4cm} I^{+}_{8} = -\int_{P'^{+}}^{\infty} \frac{dk^{+}}{P'^{+}-k^{+}},
\end{equation*}
\begin{equation*}
I^{+}_{9} = -\int_{0}^{P'^{+}} dk^{+}\frac{P'^{-}k^{+}-P'_{\perp}\cdot k_{\perp}}{k^{+}(P'^{+}-k^{+})(P'^{-}-k^{-}_{\text{on}}-(P'-k)^{-}_{\text{on}})},\hspace{.4cm} I^{+}_{10} = \int_{P'^{+}}^{\infty} dk^{+}\frac{P'^{+}}{2k^{+}(P'^{+}-k^{+})},
\end{equation*}
\begin{equation*}
I^{+}_{11} = -\int_{0}^{P'^{+}} dk^{+}\frac{P'^{+}(P'^{-}+k^{-}_{\text{on}}-(P'-k)^{-}_{\text{on}})}{2k^{+}(P'^{+}-k^{+})(P'^{-}-k^{-}_{\text{on}}-(P'-k)^{-}_{\text{on}})},\hspace{.4cm} I^{+}_{12} = -\int_{-\infty}^{0} dk^{+}\frac{P'^{+}}{2k^{+}(P'^{+}-k^{+})}.
\end{equation*}
The $P$-dependent and $P'$-dependent parts of $q_{\mu}J^{\mu}(P,P')$ in Eq. (\ref{eq:current_plus_int_sum}) separate out and are contained in the two sets of integrals $I^{+}_{1}$ through $I^{+}_{6}$, and $I^{+}_{7}$ through $I^{+}_{12}$, respectively. Substituting the sum over all $I^{+}_{i}$ in Eq. (\ref{eq:current_plus_int_sum}) eventually gives (algebra worked out in Appendix \ref{app:integrals_algebra})
\begin{equation}\label{eq:WT_current}
\begin{split}
q_{\mu}J^{\mu}(P,P') = i\int\frac{d^{2}k_{\perp}}{(2\pi)^3}\bigg[&\int_{0}^{P^{+}}dk^{+} \frac{k_{\text{on}}\cdot (P-k)_{\text{on}}+m^{2}}{k^{+}(P^{+}-k^{+})\big(P^{-}-\frac{f_{1}}{2k^{+}}-\frac{f_{2}}{2(P^{+}-k^{+})}\big)}\\
& - \int_{0}^{P'^{+}}dk^{+}\frac{k_{\text{on}}\cdot (P'-k)_{\text{on}}+m^{2}}{k^{+}(P'^{+}-k^{+})\big(P'^{-}-\frac{f_{1}}{2k^{+}}-\frac{f_{3}}{2(P'^{+}-k^{+})}\big)}\\
& - 2\int_{P^{+}}^{P'^{+}}\frac{dk^{+}}{k^{+}}\bigg].
\end{split}
\end{equation}
Thus, we see that Eq. (\ref{eq:WT_current}) is the same as Eq. (\ref{eq:WT_self-en}), and hence the Ward-Takahashi identity $q_{\mu}J^{\mu}(P,P') = \Sigma(P)-\Sigma(P')$ is established at one loop in the LF formulation.

\begin{figure}[h!]
\centering
%\subfloat[]
{\includegraphics[scale=0.65]{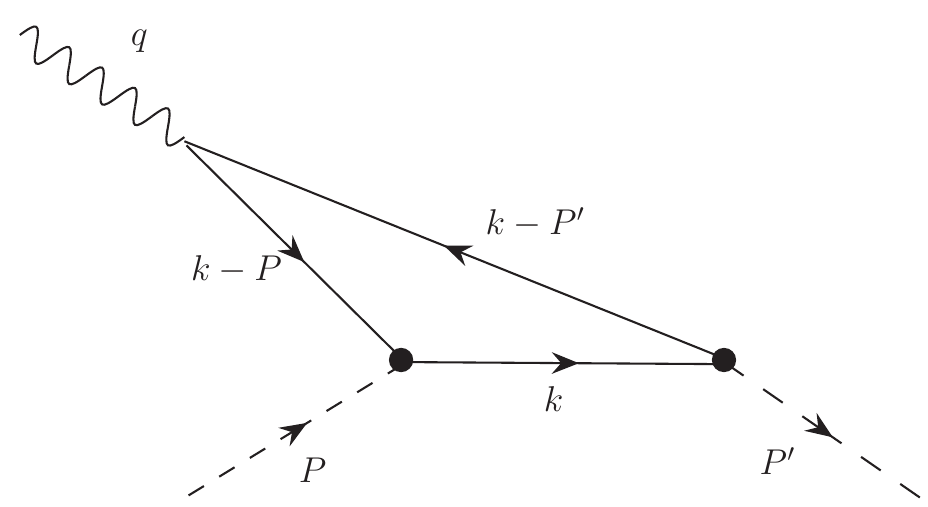}}
\caption{Pair production diagram for the electromagnetic current}
\label{fig:pair_prod}
\end{figure}

Leaving the zero-mode contribution in Eq. (\ref{eq:WT_current}) aside, it is worth noting that there are two regions of $k^{+}$ integration-one from $0$ to $P^{+}$, and another from $P^{+}$ to $P'^{+}$. Considered as a time-ordered graph in LF, the diagram in Fig. \ref{fig:triangle} can only contribute to the region $0<k^{+}<P^{+}$ due to 3-momentum conservation and the fact that the longitudinal momentum component in LF can only be non-negative. The contribution to the rest of the longitudinal momentum region, i.e., $P^{+}<k^{+}<P'^{+}$ is provided by the other time-ordered LF diagram, viz. the ``pair production'' diagram shown in Fig. \ref{fig:pair_prod}. Thus, without the pair-production diagram, the Ward-Takahashi identity in LF cannot be established.

The Ward-Takahashi Identity (WTI) is an expression of gauge invariance, and a proper regularization scheme should ideally preserve the WTI exactly. In light-front field theory, the WTI is a statement about the conserved current $J^{\mu}$. Its longitudinal structure (involving $k^{+}$) is typically what determines the existence of zero modes. A transverse ultraviolet cutoff $\Lambda_{\perp}$ (on $k_{\perp}$), provided it is independent of the LF coordinate $x^{+}$ and LF momentum $k^{+}$, is generally compatible with the structure of the light-front Ward-Takahashi identity (LFWTI). The LFWTI, being a fundamental requirement for current conservation, remains exactly satisfied under a simple transverse momentum cutoff $\Lambda_{\perp}$. The existence of zero modes is related to the integration over the longitudinal momentum $k^{+}$. The nonvalence terms are required, independent of the value of $\Lambda_{\perp}$, although their explicit functional forms will involve $\Lambda_{\perp}$.

The zero-mode contributions for the electromagnetic current, $J^{\mu}_{ZM}$, arise from the behavior of the loop integrand as $k^{+}\to 0$,
\begin{equation}
J^{\mu}_{\text {zero modes}} \sim \int d^{2}k_{\perp} [\text {Residue at } k^{+}=0].
\end{equation}

The explicit zero-mode contribution is obtained by performing the $d^{2}k_{\perp}$ integration. Therefore, the numerical magnitude of the zero-mode contribution is sensitive to $\Lambda_{\perp}$. However, regardless of the $\Lambda_{\perp}$ used, the structural role of the zero modes remains the same: it must contribute the necessary term to ensure $q_{\mu}J^{\mu}(P,P') = \Sigma(P)-\Sigma(P')$ is satisfied. The zero-mode contribution must be included for any finite $\Lambda_{\perp}$ .

In summary, the inclusion of a transverse cutoff $\Lambda_{\perp}$ does not invalidate the necessity of zero modes, but it makes the final calculated value of the current a function of this regulator, as expected.

\subsection{Electromagnetic form factor}\label{subsec:em_form_factor}

An important aspect related to the previous discussion is the calculation of electromagnetic form factors for meson bound states within the LF approach. With the calculated matrix elements of the electromagnetic current given in Eq. (\ref{eq:current}), it is possible to extract the electromagnetic elastic form factors for a bound state of two spin-half particles. This can be done using the covariant Mandelstam formula \citep{mandelstam} given by the following expression,
\begin{equation}
\langle P' | \hat{J}^{\mu} | P \rangle = (P'^{\mu}+P^{\mu})F_{M}(q^{2})
\end{equation}
and the plus or minus component of the electromagnetic current $J^{+}, J^{-}$.

However, the calculation of form factors within the LF approach needs to be carried out with care. This is because of certain symmetries like the rotational symmetry being broken, leading to contributions in addition to the valence contributions.

As established in the works by de Melo {\it et al.}, the LF projection of the electromagnetic current is often frame dependent. By comparing different current components ($J^{+}$ vs. $J^{-}$), de Melo {\it et al.} showed that nonvalence pair terms and zero modes are necessary to obtain a frame-independent form factor \citep{deMelo1999, deMelo2002}. Their models successfully describe $F_{\pi}(Q^{2})$ in both spacelike and timelike regions, incorporating vector meson dominance at the quark level \citep{deMelo2004}.

The one-loop contribution to the current in the LF frame is split into the standard (valence) term $J^{\mu}_{\text{val}}$ and the nonstandard terms $J^{\mu}_{\text{non-val}}$. The necessity of $J^{\mu}_{\text{non-val}}$ is to ensure that the LFWTI holds and that the resulting current is fully covariant. Since $J^{\mu}$ must satisfy the WTI for all values of the momentum transfer $q$, the combined contributions of the zero modes and pair-production diagrams, $J^{\mu}_{\text{non-val}}$, must explicitly depend on $q^{2}$.

Specifically, these nonstandard contributions are precisely the terms required to cancel the explicit breaking of the WTI found in $J^{\mu}_{\text{val}}$ alone, thereby restoring covariance. When decomposing the total current $\langle P' | \hat{J}^{\mu} | P \rangle$ into its form factor structure, $F(q^{2})$ will contain contributions from all sectors,
\begin{equation}
F(q^{2}) = F_{\text{val}}(q^{2}) + F_{\text{non-val}}(q^{2}).
\end{equation}

The dependence of $F_{\text{non-val}}$ on $q^{2}$ is fixed by the constraint that the final result for $F(q^{2})$ must match the covariant calculation, which, by definition, has a nontrivial dependence on $q^{2}$. Thus, the zero-mode and pair-production terms are necessarily $q^{2}$ dependent to ensure that the final result gives the correct form factor (see Refs.\citep{deMelo1999, deMelo2002, bakker2001}). It is worth noting that the zero-mode problem in the light-front approach is not restricted only to spin-half particles. For example, a good recent discussion of zero modes for a Yukawa scalar theory is given in Ref.\citep{Zhang2025}.

Within the light-front quark model, the complete electromagnetic current matrix elements are derived by combining valence and nonvalence contributions. The inclusion of the latter, which account for zero-mode effects, allows for the full restoration of covariant electromagnetic form factors.

\section{Conclusion}\label{sec:concl}

The preceding analysis delivers a rigorous, one-loop proof of the
Ward-Takahashi identity for a spinless, fermionic bound state, comparing and 
contrasting its realization within both the covariant and light-front formulations of 
quantum field theory.

In the standard covariant theory, the proof of the WT identity, which is a fundamental 
consequence of gauge invariance  and charge conservation~\cite{Zuber1980}, was established in a straightforward manner, reflecting the inherent symmetry and simplicity of the framework.

The primary technical achievement of this work lies in the successful demonstration
 of the WT identity within the more complex light-front formulation, 
 a topic for which prior demonstrations, such as the one for the boson case given by 
 Naus, de Melo, and Frederico in 1998 \cite{Naus1998}, 
 have already established the importance of careful integration techniques~\cite{deMelo1998}.

%% The primary technical achievement of this work lies in the successful
%% demonstration of the WT identity within the more %%complex Light-Front formulation, was for exemplo donne for boson case~\cite{deMelo1998}.

 We have explicitly shown that establishing this identity on the light-front is far from trivial and critically depends on properly managing the singularities introduced by the LF kinematics.

 Specifically,the proof necessitates correctly identifying and accounting for     end-point singularities     during the integration over the energy component ($k^-$)  of the loop momentum in the covariant expressions.
  These singularities, often overlooked, are the key to reconciling the LF framework with gauge invariance
  ~\cite{deMelo1997,deMelo1999}.
   Crucially, the proof requires the inclusion of both the standard regular triangle diagram for the electromagnetic current, and the pair production diagram. This finding is significant because it aligns perfectly with the physical and mathematical structure of the LF Hamiltonian formalism, where pair 
   creation, annihilation terms are essential for a complete description of the system.

By correctly incorporating these end-point singularities, the contributions typically associated with zero modes - the dynamic complexity unique to the light-front - are naturally and elegantly taken into account in the proof. 
This demonstrates that the WT identity, and thus gauge invariance, is preserved in the LF theory 
without the need for {\it ad hoc} zero-mode operators,
 provided the loop integrals are handled with scrupulous care. Along with the ``regular'' diagram viz. the triangle diagram for the electromagnetic current, the ``pair production'' diagram is also necessary to establish the proof of the identity. This is consistent with the fact that the proof is achieved in the Hamiltonian time-ordered perturbation theory on the light front.

This work not only validates the light-front hamiltonian time-ordered perturbation theory as 
a consistent framework for handling gauge theories but also provides a clear, instructive procedure for translating between covariant and LF results. The explicit connection made between end-point singularities in loop integrals and the required physical contributions (like pair production diagrams) significantly enhances our understanding of how fundamental symmetries, like     gauge invariance, are preserved in nontrivial field-theoretic contexts within the light-front framework. This rigorous verification strengthens the theoretical 
foundation for future applications of light-front quantization to more realistic and 
complex bound-state problems in particle and nuclear physics.

\section*{Acknowledgments}
D. B. acknowledges the support from Conselho Nacional de Desenvolvimento Científico e Tecnológico - CNPq, Brazil, Process No. 152348/2024-7. J. P. B. C. M. acknowledges the support from Funda\c{c}\~{a}o de Amparo \`{a} Pesquisa do Estado de S\~{a}o Paulo~(FAPESP), 
Process No. 2023/09539-1,
Instituto Nacional de Ci\^{e}ncia e Tecnologia-Nuclear Physics and Applications~(INCT-FNA, MCTI), Brazil,
Process No. 464898/2014-5, 
and FAPESP, Brazil,  thematic Project, No. 2024/17816-8.

\appendix 

\section{} \label{app:conventions}
The following conventions are used for the LF components of the coordinate four-vector $x^{\mu}$ in the 
present work: 
$x^{+}=\frac{x^{0}+x^{3}}{\sqrt{2}}$~(time coordinate),
$x^{-}=\frac{x^{0}-x^{3}}{\sqrt{2}}$~(longitudinal space coordinate)
and ${\bf{x}}^{\perp}=(x^{1}, x^{2})$ (transverse space coordinates).

The conjugate LF components of the momentum four-vector $p^{\mu}$ are~$p^{-}=\frac{p^{0}-p^{3}}{\sqrt{2}}$~(energy),~$p^{+}=\frac{p^{0}+p^{3}}{\sqrt{2}}$ 
(longitudinal momentum), and ${\bf{p}}^{\perp}=(p^{1}, p^{2})$~(transverse momentum).

The metric tensor in this convention is\\
$g_{\mu\nu}=g^{\mu\nu}=$
$ \begin{bmatrix}
0 & 1 & 0 & 0 \\
1 & 0 & 0 & 0 \\
0 & 0 & -1 & 0 \\
0 & 0 & 0 & -1 
\end{bmatrix} $.\\

The $\gamma$ matrices are given by the following representation
\begin{equation}
\gamma^{+}=\frac{\gamma^{0}+\gamma^{3}}{\sqrt{2}},\hspace{.5cm} \gamma^{-}=\frac{\gamma^{0}-\gamma^{3}}{\sqrt{2}},
\end{equation}
where
\begin{equation}
\gamma^{0}=\begin{bmatrix}
0 & I \\
I & 0 \\
\end{bmatrix},\hspace{.5cm} \gamma^{k}=\begin{bmatrix}
0 & -\sigma_{k} \\
\sigma_{k} & 0 \\
\end{bmatrix}.
\end{equation}

The $\gamma$ matrices satisfy the relations
\begin{equation}
 \begin{split}
 &\{\gamma^{\alpha}, \gamma^{\beta}\}=2g^{\alpha\beta}\mathds{1}, \\ &\{\gamma^{5},\gamma^{\mu}\} = 0,\\ &(\gamma^{+})^{2}=(\gamma^{-})^{2}=0.
 \end{split}
\end{equation}

\section{} \label{app:integrals_algebra}

Consider the $P$-dependent set of integrals in Eq. (\ref{eq:current_plus_int_sum}). $I^{+}_{1}$ can be written as
\begin{equation*}
\begin{split}
I^{+}_{1} &= \int_{0}^{\infty} \frac{dk^{+}}{P^{+}-k^{+}} - \int_{0}^{P^{+}} \frac{dk^{+}}{P^{+}-k^{+}}\\
&= \int_{-\infty}^{P^{+}} \frac{dk^{+}}{k^{+}} - \int_{0}^{P^{+}} dk^{+}\frac{k^{+}(P^{-}-k^{-}_{\text{on}}-(P-k)^{-}_{\text{on}})}{k^{+}(P^{+}-k^{+})(P^{-}-k^{-}_{\text{on}}-(P-k)^{-}_{\text{on}})}.
\end{split}
\end{equation*}
$I^{+}_{2}$ can be written as
\begin{equation*}
\begin{split}
I^{+}_{2} &= -\int_{-\infty}^{0} \frac{dk^{+}}{P^{+}-k^{+}} - \int_{0}^{P^{+}} \frac{dk^{+}}{P^{+}-k^{+}}\\
&= -\int_{P^{+}}^{\infty} \frac{dk^{+}}{k^{+}} - \int_{0}^{P^{+}} \frac{dk^{+}}{k^{+}}.
\end{split}
\end{equation*}
$I^{+}_{6}$ can be written as
\begin{equation*}
\begin{split}
I^{+}_{6} = -\int_{P^{+}}^{\infty} dk^{+}\frac{P^{+}}{2k^{+}(P^{+}-k^{+})} &= -\int_{P^{+}}^{\infty} \frac{dk^{+}}{2k^{+}} - \int_{P^{+}}^{\infty} \frac{dk^{+}}{2(P^{+}-k^{+})}\\
&= -\int_{P^{+}}^{\infty} \frac{dk^{+}}{2k^{+}} - \int_{-\infty}^{0} \frac{dk^{+}}{2k^{+}}.
\end{split}
\end{equation*}
Also, the integrals $I^{+}_{3}$, $I^{+}_{4}$ and $I^{+}_{5}$ add up to give
\begin{equation*}
\begin{split}
I^{+}_{3}+I^{+}_{4}+I^{+}_{5} &= \int_{0}^{P^{+}} dk^{+}\frac{P^{+}k^{-}_{\text{on}}+P^{-}k^{+}-P_{\perp}\cdot k_{\perp}}{k^{+}(P^{+}-k^{+})(P^{-}-k^{-}_{\text{on}}-(P-k)^{-}_{\text{on}})} + \int_{-\infty}^{P^{+}} dk^{+}\frac{P^{+}}{2k^{+}(P^{+}-k^{+})}\\
&= \int_{0}^{P^{+}} dk^{+}\frac{P^{+}k^{-}_{\text{on}}+P^{-}k^{+}-P_{\perp}\cdot k_{\perp}}{k^{+}(P^{+}-k^{+})(P^{-}-k^{-}_{\text{on}}-(P-k)^{-}_{\text{on}})} + \int_{-\infty}^{P^{+}} \frac{dk^{+}}{2k^{+}} + \int_{-\infty}^{P^{+}} \frac{dk^{+}}{2(P^{+}-k^{+})}\\
&= \int_{0}^{P^{+}} dk^{+}\frac{P^{+}k^{-}_{\text{on}}+P^{-}k^{+}-P_{\perp}\cdot k_{\perp}}{k^{+}(P^{+}-k^{+})(P^{-}-k^{-}_{\text{on}}-(P-k)^{-}_{\text{on}})} + \int_{-\infty}^{P^{+}} \frac{dk^{+}}{2k^{+}} + \int_{0}^{\infty} \frac{dk^{+}}{2k^{+}}\\
&= \int_{0}^{P^{+}} dk^{+}\frac{P^{+}k^{-}_{\text{on}}+P^{-}k^{+}-P_{\perp}\cdot k_{\perp}}{k^{+}(P^{+}-k^{+})(P^{-}-k^{-}_{\text{on}}-(P-k)^{-}_{\text{on}})} + \int_{-\infty}^{0} \frac{dk^{+}}{2k^{+}} + \int_{0}^{P^{+}} \frac{dk^{+}}{k^{+}} + \int_{P^{+}}^{\infty} \frac{dk^{+}}{2k^{+}}.
\end{split}
\end{equation*}
We see that the second term of $I^{+}_{2}$, the two terms of $I^{+}_{6}$, and the last three terms of $(I^{+}_{3}+I^{+}_{4}+I^{+}_{5})$ cancel between themselves. Also, the numerator of the second term of $I^{+}_{1}$ and of the first term of $(I^{+}_{3}+I^{+}_{4}+I^{+}_{5})$ can be combined as follows.
\begin{equation*}
\begin{split}
&k^{+}P^{-}+k^{-}_{\text{on}}P^{+}-k_{\perp}\cdot P_{\perp} - k^{+}(P^{-}-k^{-}_{\text{on}}-(P-k)^{-}_{\text{on}})\\
&= k^{+}P^{-}+k^{-}_{\text{on}}P^{+}-k_{\perp}\cdot P_{\perp} - k^{+}(P^{-}-k^{-}_{\text{on}}-(P-k)^{-}_{\text{on}}) - 2k^{+}k^{-}_{\text{on}} + 2k^{+}k^{-}_{\text{on}}\\
&= k^{+}P^{-}+k^{-}_{\text{on}}P^{+}-k_{\perp}\cdot P_{\perp} - k^{+}(P^{-}+k^{-}_{\text{on}}-(P-k)^{-}_{\text{on}}) + k^{2}_{\perp} + m^{2}\\
&= k^{+}(P-k)^{-}_{\text{on}} + k^{-}_{\text{on}}(P^{+}-k^{+}) - k_{\perp}\cdot (P_{\perp}-k_{\perp}) + m^{2}\\
&= k_{\text{on}}\cdot (P-k)_{\text{on}}+m^{2}.
\end{split}
\end{equation*}
Thus, the $P$-dependent integrals of $q_{\mu}J^{\mu}(P,P')$ in Eq. (\ref{eq:current_plus_int_sum}) give
\begin{equation}\label{I1-I6 sum}
\begin{split}
\sum_{i=1}^{6} I^{+}_{i} &= \int_{0}^{P^{+}}dk^{+} \frac{k_{\text{on}}\cdot (P-k)_{\text{on}}+m^{2}}{k^{+}(P^{+}-k^{+})\big(P^{-}-\frac{f_{1}}{2k^{+}}-\frac{f_{2}}{2(P^{+}-k^{+})}\big)} + \int_{-\infty}^{P^{+}} \frac{dk^{+}}{k^{+}} - \int_{P^{+}}^{\infty} \frac{dk^{+}}{k^{+}}\\
&= \int_{0}^{P^{+}}dk^{+} \frac{k_{\text{on}}\cdot (P-k)_{\text{on}}+m^{2}}{k^{+}(P^{+}-k^{+})\big(P^{-}-\frac{f_{1}}{2k^{+}}-\frac{f_{2}}{2(P^{+}-k^{+})}\big)} + \int_{-\infty}^{P^{+}} \frac{dk^{+}}{k^{+}} - \int_{P^{+}}^{P'^{+}} \frac{dk^{+}}{k^{+}} - \int_{P'^{+}}^{\infty} \frac{dk^{+}}{k^{+}}.
\end{split}
\end{equation}

Similarly, the $P'$-dependent contribution to $q_{\mu}J^{\mu}(P,P')$ is obtained from $I^{+}_{7}$ through $I^{+}_{12}$, and is given by
\begin{equation}\label{eq:I7-I12 sum}
\begin{split}
\sum_{i=7}^{12} I^{+}_{i} &= -\int_{0}^{P'^{+}}dk^{+} \frac{k_{\text{on}}\cdot (P'-k)_{\text{on}}+m^{2}}{k^{+}(P'^{+}-k^{+})\big(P'^{-}-\frac{f_{1}}{2k^{+}}-\frac{f_{3}}{2(P'^{+}-k^{+})}\big)} - \int_{-\infty}^{P'^{+}} \frac{dk^{+}}{k^{+}} + \int_{P'^{+}}^{\infty} \frac{dk^{+}}{k^{+}}\\
& = -\int_{0}^{P'^{+}}dk^{+} \frac{k_{\text{on}}\cdot (P'-k)_{\text{on}}+m^{2}}{k^{+}(P'^{+}-k^{+})\big(P'^{-}-\frac{f_{1}}{2k^{+}}-\frac{f_{3}}{2(P'^{+}-k^{+})}\big)} - \int_{-\infty}^{P^{+}} \frac{dk^{+}}{k^{+}} - \int_{P^{+}}^{P'^{+}} \frac{dk^{+}}{k^{+}} + \int_{P'^{+}}^{\infty} \frac{dk^{+}}{k^{+}}.
\end{split}
\end{equation}

Equations (\ref{I1-I6 sum}) and (\ref{eq:I7-I12 sum}) add up to give

\begin{equation}
\begin{split}
\sum_{i=1}^{12} I^{+}_{i} = &\int_{0}^{P^{+}}dk^{+} \frac{k_{\text{on}}\cdot (P-k)_{\text{on}}+m^{2}}{k^{+}(P^{+}-k^{+})\big(P^{-}-\frac{f_{1}}{2k^{+}}-\frac{f_{2}}{2(P^{+}-k^{+})}\big)}\\
& - \int_{0}^{P'^{+}}dk^{+} \frac{k_{\text{on}}\cdot (P'-k)_{\text{on}}+m^{2}}{k^{+}(P'^{+}-k^{+})\big(P'^{-}-\frac{f_{1}}{2k^{+}}-\frac{f_{3}}{2(P'^{+}-k^{+})}\big)}\\
& - 2\int_{P^{+}}^{P'^{+}} \frac{dk^{+}}{k^{+}}.
\end{split}
\end{equation}

\bibliography{references.bib}

\end{document}